\documentclass[copyright,creativecommons]{eptcs}
\providecommand{\event}{WWV 2011} 
\usepackage{breakurl}             
\usepackage[noadjust]{cite}
\usepackage{graphicx,url,psfrag}
\usepackage{mathptm}
\usepackage{amsmath,amssymb,amsfonts,amsthm,latexsym,amstext}
\usepackage{verbatim} 
\usepackage{color,times}
\usepackage{wasysym}
\usepackage{txfonts}
\newtheorem{thm}{Theorem}[section]

\newtheorem{defn}[thm]{Definition}

\newcommand{\Seq}[3]{(#1>\!\! #2\!\! >#3)}

\newcommand{\Par}[3]{(#1<\!\! #2\!\! <#3)}

\newcommand{\Ind}[2]{(#1 \mid #2)}

\title{Product Lines for Service Oriented Applications - PL for SOA}
\author{Maurice H. ter Beek \qquad\qquad\quad Stefania Gnesi
\institute{Istituto di Scienza e Tecnologie dell'Informazione, ISTI--CNR\\
Pisa, Italy}
\email{\{maurice.terbeek,stefania.gnesi\}@isti.cnr.it}
\and Mercy N. Njima
\institute{IMT Institute for Advanced Studies Lucca\\ Lucca, Italy}
\email{mercy.njima@imtlucca.it}
}
\def\titlerunning{PL for SOA}
\def\authorrunning{M.H. ter Beek, S. Gnesi \& M.N. Njima}

\begin{document}
\maketitle

\begin{abstract}
PL for SOA proposes, formally, a software engineering methodology, development 
techniques and support tools for the provision of service product lines. We 
propose rigorous modeling techniques for the specification and 
verification of formal notations and languages for service computing with 
inclinations of variability. Through these cutting-edge technologies, increased levels of 
flexibility and adaptivity can be achieved. This will involve developing semantics of 
variability over behavioural models of services. Such tools will assist organizations 
to plan, optimize and control the quality of 
software service provision, both at design and at run time by making it possible to develop flexible and 
cost-effective software systems that support high levels of reuse. We tackle this challenge from 
two levels. We use feature modeling from product line engineering and, from a services 
point of view, the orchestration language Orc. We introduce the Smart Grid as the service 
product line to apply the techniques to.
\end{abstract}

\section{Introduction}
Business environments command innovation, increasingly shorter 
time-to-market and efficiency. Product line technology, 
is increasingly finding its way to the software sector, allowing companies to sustain 
growth and achieve market success~\cite{BG08}. 

Service-Oriented Architecture (SOA) has emerged as a standard-based computing model for designing, 
building and deploying flexible distributed software applications. SOA emphasizes extremely loosely-coupled design approaches where disparate systems with different computing platforms can collaborate 
and evolve without major changes to their core architectures. Services are designed as self-contained 
modules that can be advertized, discovered, composed and negotiated on demand.

Software Product Lines (SPL) are families of software systems that share common functionality, but 
each member also has variable functionality. The main goal of SPL is the agile and speedy development of 
member systems by taking advantage of reusable assets from all phases of the development life cycle. 
This goal is similar to SOA's goals~\cite{IBM}.

Despite the wide academic and industrial activities related to SOA, no systematic end-to-end 
methodology exists to analyze and design service-oriented applications. On the other hand, SPL 
is an established field with considerable methodological support. It is then clear that combining 
SOA and SPL is a powerful way to build complex evolving systems. 

The combination of SPL and SOA development practices is a new development 
paradigm that can help provide the answers to the need for agility, 
versatility and economy. SOA and SPL approaches to software 
development share a common goal. They both encourage an organization to reuse existing assets
and capabilities rather than repeatedly redeveloping them for new systems. These approaches 
enable organizations to capitalize on reuse to achieve desired benefits such as productivity gains,
decreased development costs, improved time to market, higher reliability and competitive 
advantage. Their distinct goals may be stated as follows~\cite{SOAPL07,SOAPL08, SOAPL09}:
\begin{description}
\item[SOA] To enable the assembly, orchestration and maintenance of enterprise solutions to quickly
      react to changing business requirements. 
\item[SPL] To systematically capture and exploit commonality among a set of related systems while
      managing variations for specific customers or market segments. 
\end{description}
The goal of our research is to formally answer the question, \lq How can the use of product line 
practices support service-oriented applications?\rq\ SOA and SPL have their differences and 
similarities and we are exploiting the similarities in order 
to naturally formalize SPL~\cite{Gruler} and service-oriented applications~\cite{PT07}. 

This work is part of a longer-term research effort to develop a PL for SOA formalisms.
The paper introduces a first step towards this goal. We have found an existing formalism 
for modeling variability in product 
lines that can be combined with a SOA calculus and thus create a PL for SOA formalism. 

Modeling variability in product families has been the subject of extensive study 
in the literature on SPL, especially that concerning feature 
modeling~\cite{Batory,Czarnecki,FODA}.
Variability modeling addresses how to define which features or components of a 
system are optional, alternative, mandatory, required or excluded; formal methods are then developed 
to show that a product belongs to a family, or to derive instead a product from a 
family, by means of a proper selection of the features or components. Variability management is the key aspect that differentiates SPL engineering from \lq conventional\rq\ software engineering.

Labelled Transition Systems (LTSs) have been used successfully to reason
about system behaviour. Modal Transition Systems (MTSs) are an extension
of LTSs that distinguish between mandatory, possible and unknown behaviour.
MTSs have been studied for some time as a means for formally describing partial
knowledge of the intended behaviour of software systems~\cite{20years}.

MTSs have been proposed as a formal model for product 
families~\cite{Uchitel,Larsen07}, allowing one to embed in a single model the 
behaviour of a family of products that share the basic structure of states and 
transitions, transitions which can moreover be seen as mandatory or possible for 
the products of the family. 
In~\cite{SPLC08}, the MTS concept was pushed to a more general form, allowing 
more precise modeling of the different kinds of variability that can typically be 
found in the definition of a product family.

In~\cite{ABFG10b}, a temporal logic for modeling variability in product families was proposed by taking advantage of the way in which deontic logic formalises concepts like violation, obligation, permission and prohibition, using MTSs as the underlying semantic model. 
In~\cite{ABFG11a,ABF+11}, a model checker is presented based on a formal framework consisting of {\sc vaCTL} (a variability and action-based branching-time temporal logic) with its natural interpretation structure (MTSs). Product derivation is defined inside the framework and logical formulae are used as variability constraints as well as behavioural properties to be verified for families and products alike. A first attempt to apply this tool to the analysis of variability in behavioural descriptions of families of services is presented in~\cite{ABFG11b}.

From a service orientation point of view, we believe that Cook and Misra's Orc~\cite{web} will 
highlight the SPL aspects necessary to meet our goals. An operational semantics of Orc based on LTSs appears in~\cite{Timed}. Thus, it appears from a first glance that we can extend the LTS semantics of Orc to a semantics over MTSs and merge the different scopes of SOA and SPL creating a PL for SOA formalism. The formal definition of such a semantics is left for future work.

The paper is structured as follows. In Section~\ref{Related}, we review some related work and 
some product line basics from a feature 
modeling perspective and present the orchestration language Orc and its usefulness. 
Section~\ref{SOPL} combines ideas from product line engineering and the service-oriented concurrency calculus 
of Orc. Section~\ref{SM} highlights the case study over which the tools are applied. We conclude in 
Section~\ref{FW} with some remarks on future work.

\section{Related Work and Preliminaries}\label{Related}
Various authors have contributed to the study of combining SOA and SPL practices and most are still 
in the preliminary stages of defining what SPL for SOA means: \cite{IBM} presents a 
method to design service-oriented applications based on SPL principles by applying variability analysis techniques to Web Services to design customized service-based applications.

In~\cite{Tension,Wienands} the authors study the common problems relating to SOA and SPL approaches 
and propose ways of reconciling the two, while~\cite{Asadi} demonstrates how model-driven engineering 
can help with injecting a set of required commonalities
and variability of a software product from a high-level business
process design to the lower levels of service use. To realize the method and activities involved, a 
supply chain management application is used.

G{\"u}nther et al.~\cite{GuntherB08} propose a differentiated
development process for SPLs implementing a SOA. They use an 
extensive example of a web store to show how parts of this process can be solved technically with already 
developed methods for feature modeling and management using Web Services.

An approach to service identification methods is proposed in~\cite{Baik} to bridge the feature
models of product lines and the business process models in service orientation and enables functions 
to be expressed as services.

\subsection{Product Line Modeling}\label{PL}
As a first step we have used feature diagrams to model product lines. Feature diagrams
are a family of popular modeling languages used for engineering requirements in SPL 
represented as the nodes of a tree, with the product family being the root and having the following features~\cite{FODA}:
\begin{itemize}
\item \emph{optional\/} features, may be present in a product only if their parent is present;
\item \emph{mandatory\/} features, are present in a product if and only if their parent is present;
\item \emph{alternative\/} features, are a set of features among which one and only one is present in a 
product if their parent is present. 
\end{itemize}
When additional constraints are added to a feature diagram, this results in a feature model. 
Constraints come in several flavours and we consider the following constraints:
\begin{itemize}
\item \textit{requires\/} is a unidirectional relation between two features indicating that the presence of one 
feature requires the presence of the other;
\item \textit{excludes\/} is a bidirectional relation between two features indicating that the presence of either 
feature is incompatible with the presence of the other.
\end{itemize}

\subsection{Service-Oriented Modeling}\label{SO}
Orc has proved to be a high-level language that 
makes the notoriously difficult task of building distributed systems easier. It coordinates interactions 
among basic subsystems, called sites, by use of a small number of combinators.
It allows integration of components and assumes that structured concurrent programs should 
be developed much like structured sequential programs, by decomposing a problem and combining the 
solutions with the combinators.

Orc permits structuring programs in a hierarchical manner, while permitting interactions among 
subsystems in a controlled 
way~\cite{task-orch}. The basis for its design is to allow integration of components and is founded 
on the premise of combination. Thus, combinators are a very important part of the theory.

An Orc program consists of a goal expression (either primitive or a combination of two expressions) 
and a set of definitions. The goal expression is evaluated in order to run the program. The definitions 
are used in the goal and in other definitions. A component is generally called a service; Orc adopts 
the more neutral term \textit{site\/} which is the most primitive Orc expression. It represents an external 
program 
and is said to publish a value when a value is returned in response to a call.

Given the formalism we are working on, we will not dwell on the Orc programming language but concentrate 
on the Orc calculus.

\subsection{The Orc Calculus}
We present the calculus informally in this paper. 
The Orc calculus is based on the execution of expressions. Expressions are built
up recursively using Orc's concurrent combinators~\cite{Prog}. When executed, an Orc expression 
calls services and may publish values. Different executions of the same
expression may have completely different behaviour; they may call different services, 
receive different responses from the same service and publish different values.
Orc expressions use sites to refer to external services. A site may be implemented 
on the client's machine or on a remote machine. A site may provide any
service; it could run sequential code, transform data, communicate with a Web
Service or be a proxy for interaction with a human user.

A site call is defined as $A(x)$, where $A$ is a site name and $x$ is a list of actual inputs. The following 
table lists the fundamental sites of Orc.\\

\begin{center}
\begin{tabular}{ll}
$\textit{if}(b)$: & Returns a value if $b$ is true, and otherwise does not respond. \\ 
$\textit{Rtimer}(t)$: & Returns a value after exactly $t$, $t\geqslant 0$, time units. \\ 
$\textit{Signal}()$: & Returns a value immediately. Same as $\textit{if}(\textit{true})$. \\ 
$0$: & Blocks forever. Same as $\textit{if}(\textit{false})$.\\
\end{tabular}
\end{center}

Though the Orc calculus itself contains no sites, for our purposes we consider another fundamental
site which is essential to writing useful computations. The site \textit{let\/} is the 
identity site; when passed one argument,
it publishes that argument, and when passed multiple arguments it publishes
them as a tuple. 

Orc has four combinators to compose expressions: the parallel combinator $\mid$,
the sequential combinator $>x>$, the asymmetric parallel combinator $<x<$, and the otherwise 
combinator $;$. When composing expressions, the $>x>$ combinator has the highest precedence, 
followed by $\mid$, then $<x<$ and finally $;$ has the lowest precedence~\cite{Prog}. 

\begin{enumerate}
\item The independent parallel combinator, $\Ind{A}{B}$, allows independent 
concurrent execution of $A$ and $B$. The sites called by $A$ and $B$ individually 
are called by $\Ind{A}{B}$ and the values published by $A$
and $B$ are published by $\Ind{A}{B}$.
\item The sequential combinator, $\Seq{A}{x}{B}$, 
initiates a new instance of $B$ for every value 
published by $A$ whose value is bound to name $x$ in that instance of $B$. 
The values published by $\Seq{A}{x}{B}$ are all instances of those published 
by $B$. If $x$ is not used in $B$, this combinator is abbreviated by $(A \gg B)$.
\item The asymmetric parallel combinator, $\Par{A}{x}{B}$, 
evaluates $A$ and $B$ independently, 
but the site calls in $A$ that depend on $x$ are suspended
until $x$ is bound to a value; 
the first value from $B$ is bound to $x$, evaluation of $B$
is then terminated and suspended calls in $A$ are resumed; the values published
by $A$ are those published by $\Par{A}{x}{B}$. If $x$ is not used in $B$, this combinator is abbreviated by $(A \ll B)$.
\item The otherwise combinator, $(A;B)$, executes $A$ and if it 
completes and has not published any values, then executes $B$. 
If $A$ did publish one or more values, then $B$ is ignored. 
The publications of $(A;B)$ are thus those of $A$ if $A$ publishes, 
or those of $B$ otherwise~\cite{task-orch}. 
\end{enumerate}

\subsection{Modal Transition Systems} \label{Semantics}
MTSs are now an accepted formal model for defining behavioural aspects of product families~\cite{20years,ABFG10b,ABFG11a,ABFG11b,ABF+11,SPLC08,Uchitel,Larsen07}. An MTS is an LTS with a distinction between may and must transitions, seen as \emph{optional\/} or \emph{mandatory\/} features for a family's products. For a given product family, an MTS can model
\begin{itemize}
\item its \emph{underlying behaviour\/}, shared among all products, and 
\item its \emph{variation points\/}, differentiating between products.
\end{itemize}
An MTS cannot model advanced variability constraints regarding {\it alternative\/} features nor those regarding the {\it requires\/} and {\it excludes\/} inter-feature relations~\cite{ABFG10b}. Such advanced variability constraints can be formalized by means of an associated set of logical formulae expressed in the variability and action-based branching-time temporal logic {\sc vaCTL} (interpreted over MTSs)~\cite{ABFG11a}. 

We now formally define MTSs and --- to begin with --- their underlying LTSs.

\begin{defn}
An LTS is a quadruple $(Q, A, \overline{q}, \delta)$, with set $Q$ of states, set $A$ of actions, initial state $\overline{q}\in Q$, and transition relation $\delta\subseteq Q\times A\times Q$.

One may also write $q\xrightarrow{a} q'$ for $(q,a,q')\in\delta$.
\label{def:LTS}
\qed\end{defn}

In an MTS, transitions are defined to be possible (\emph{may\/}) or mandatory (\emph{must\/}).

\begin{defn}
An MTS is a quintuple $(Q, A, \overline{q}, \delta^\Box, \delta^\Diamond)$ such that the quadruple $(Q, A, \overline{q}, \delta^\Box\cup\delta^\Diamond)$ is an LTS, called its \emph{underlying\/} LTS. 
An MTS has two transition relations: $\delta^\Diamond\subseteq Q\times A\times Q$ is the \emph{may\/} transition relation, expressing \emph{possible\/} transitions, while $\delta^\Box\subseteq Q\times A\times Q$ is the \emph{must\/} transition relation, expressing \emph{mandatory\/} transitions. 
By definition, $\delta^\Box\subseteq\delta^\Diamond$. 

One may also write $q\xrightarrow{a}_\Box q'$ for $(q,a,q')\in\delta^\Box$ and $q\xrightarrow{a}_\Diamond\!q'$ for $(q,a,q')\in\delta^ \Diamond$.  
\label{def:MTS}
\qed\end{defn}

The inclusion $\delta^\Box\subseteq\delta^\Diamond$ formalises that mandatory transitions must also be possible. Reasoning on the existence of transitions is thus like reasoning with a 3-valued logic with the truth values \emph{true\/}, \emph{false\/}, and \emph{unknown\/}: mandatory transitions ($\delta^\Box$) are \emph{true\/}, possible but not mandatory transitions ($\delta^\Diamond\setminus\delta^\Box$) are \emph{unknown\/}, and impossible transitions ($(q,a,q')\notin\delta^\Box\cup\delta^\Diamond$) are \emph{false\/}.

To model feature model representations of product families as MTSs one thus needs a \lq translation\rq\ from features to actions (not necessarily a one-to-one mapping) and the introduction of a behavioural relation (temporal ordering) among them. A family's products are then considered to differ w.r.t.~the actions they are able to perform in any given state of the MTS. This means that the MTS of a product family has to accommodate all the possibilities desired for each derivable product, predicating on the choices that make a product belong to that family. 

Figure~\ref{fig:Model1} below is an example of an MTS: dashed arcs are used for the may transitions that are not must transitions ($\delta^\Diamond\setminus\delta^\Box$) and solid ones for must transitions ($\delta^\Box$). 

Given an MTS description of a product family, an LTS describing a product is obtained by preserving at least all must transitions and turning some of the may transitions (that are not must transitions) into must transitions as well as removing all of the remaining may transitions.

\begin{defn}
Let ${\cal F} = (Q, A, \overline{q}, \delta^\Box, \delta^\Diamond)$ be an MTS specifying a product family.
A set of \emph{products\/} specified as a set of LTSs $\{\,{\cal P}_i = (Q_i, A, \overline{q}, \delta_i)\mid i > 0\,\}$ is derived by considering each transition relation $\delta_i$ to be $\delta^\Box\cup\,R$, with $R\subseteq\delta^\Diamond$, defined over a set of states $Q_i\subseteq Q$, so that $\overline{q}\in Q_i$, and every $q\in Q_i$ is reachable from $ \overline{q}$ via transitions from $\delta_i$.

More precisely, we say that ${\cal P}_i$ \emph{is a product of\/} ${\cal F}$, denoted by ${\cal P}_i\vdash {\cal F}$, if and only if $\overline{q}_i\vdash\overline{q}$, where $q_i\vdash q$ holds, for some $q_i\in Q_i$ and $q\in Q$, if and only if:
\begin{itemize}
\item whenever $q\xrightarrow{a}_{\Box} q'$, for some $q'\in Q$, then $\exists\,q'_i\in Q_i: q_i\!\xrightarrow{a}_i q'_i$ and $q'_i\vdash q'$, and
\item whenever $q_i\xrightarrow{a}_i q'_i$, for some $q'_i\in Q_i$, then $\exists\,q' \in Q:$ $q \xrightarrow{a}_{\Diamond} q'$ and $q'_i\vdash q'$.
\qed\end{itemize}
\label{def:MTS product derivation}
\end{defn}
The products derived in this way obviously might not satisfy the aforementioned advanced variability constraints that MTSs cannot model. 
However, as said before, {\sc vaCTL} can be used to express those constraints and~\cite{ABFG11b} contains an algorithm to derive from an MTS all products that are valid w.r.t.~constraints expressed in {\sc vaCTL}.

\section{Service-Oriented Product Line}\label{SOPL}
In order to model a service product line, we merge the feature modeling and Orc approaches. We show here 
that the Orc calculus can be viewed from product line/feature modeling perspective and, hence, the resulting 
calculus can sufficiently specify service-oriented product lines. We believe that the Orc combinators can be given a semantics over MTSs that would result in an almost one-to-one correspondence with the features and inter-feature relations of product families:
\begin{itemize}
\item The independent parallel combinator, $\Ind{A}{B}$, can be used to specify mandatory features. This is 
because there is no direct communication or interaction between these two computations and they are instantiated 
independently and in parallel.
\item The sequential combinator, $\Seq{A}{x}{B}$, can be used to specify required features. This follows from
the fact that $B$ is never instantiated unless $A$ publishes a value which is bound to $x$ and utilized as 
input in $B$. Thus, $B$ requires published values from $A$.
\item The asymmetric parallel combinator, $\Par{A}{x}{B}$, can specify optional features. Since both $A$ and $B$ are instantiated in parallel and those computations of $A$ that require a value from $B$ are suspended, this combinator may ignore the published value from $B$ in order to incorporate optionality. 
\item The otherwise combinator, $(A;B)$, can be used to specify excluding features especially when there is a 
preferred outcome or priority. It follows because the computation of either $A$ or $B$ means that the other 
cannot be instantiated or has already failed.
\end{itemize}
We do not see how to directly cater for the alternative features from the combinators. However, we foresee the use of Orc's powerful composition of the combinators to reason about them. We then intend to look at an alternative feature as a choice between two computations from which we let only one proceed. This is the essence of mutual exclusion. We consider a product in which we choose feature $M$ if $A$ happens, while otherwise we choose $N$ (i.e.~if $B$ happens). We represent $A$ and $B$ as sites and $M$ and $N$ as expressions and use site \textit{flag\/} to record which of $A$ 
and $B$ responds first.
$$\textit{if}(\textit{flag})\gg M \mid \textit{if}(\neg \textit{flag}) \gg N \ll \textit{flag} \in (A \gg \textit{let}(\textit{true})) \mid (B \gg \textit{let}(\textit{false}))$$
In the future, we plan to extend the LTS semantics of Orc to a semantics over MTSs to merge the different scopes of SOA and SPL and create a PL for SOA formalism. 

\section{Case Study}\label{SM}
The energy utilities industry may be one of the last great technological frontiers, due to 
the fact that it has experienced little innovation over its lifespan and it is quickly approaching the 
end of its design life.

However, the utility industry is about to embark on a revolutionary journey: the Smart Grid. Utilities and 
information technology companies will be surrounding the electric grid with a digital grid that 
will provide consumers and businesses with many value propositions~\cite{Cisco}.

One of the key components to this \lq smart\rq\ electric grid is the upgrade to a two-way 
communications technology. This technology, partially fueled by governments supporting the modernization 
of the electric grid, requires one of the largest IT \lq upgrades\rq\ that we will see in decades, and 
provides new product, service and market opportunities for utilities, generators, power traders and 
information technology companies~\cite{Cisco1}.

We have a long way to go to turn this antiquated grid into a Smart Grid. However, using SOAs which lead to decreasing the time to market we may be able to have a fast response to 
the market needs. The applications will entail a fast time-to-market response, 
correctness, reusability, maintainability, testability and evolvability --- besides low cost. 

Furthermore, like the Internet it will require a standard layered and distributed 
architecture in order to deliver electricity over a two-way protocol from supplier to consumer 
utilizing independent components that must cooperate. SPLs and SOAs can provide several of these requirements due to the inherent 
flexibility in composing more sophisticated complex systems.

Suppose that your utility company has developed an intelligent 
electrical power system that 
leverages increased use of communications and information technology in the generation, delivery 
and consumption of electrical energy. Your company provides a choice among a family of products 
with different price tags and different functionalities. The basic architecture provides three products: 
\begin{enumerate}
\item \textit{Integration of renewables\/}, offering storage 
capacity, vehicle to grid and electric vehicles.
\item \textit{Demand response\/}, offering efficient markets, 
load shifting and incorporating 
all end users.
\item \textit{Grid monitoring management\/}, offering smart meters, self-healing capability 
and integrated 
communications.
\end{enumerate}
The coordination component uses predefined external services (one for 
each business sector) to retrieve a list of alternatives, say for storage, load shifting or billing 
through the smart meters.

The basic product can be enhanced in two ways:
\begin{enumerate}
\item Adding the possibility to choose what company to source your electricity
and in case you have generation capacity, choosing whom you will sell your excess power to, 
from a set of utilities in order to retrieve the best quotes through more than one service. 
\item Adding the possibility for the user to make a reservation for the supply of extra electricity. 
This is accomplished by means of an added component that requests an external forecasting service 
to predict what 
sources of electricity generation are available and how much demand exists. 
\end{enumerate}
These enhancements can be combined to obtain four different products of the family. A greater 
level of flexibility in the service may be added by incorporating dynamic roles.

As the grid becomes 
more intelligent and more complex, the tools to operate it become increasingly important.
Hence the need for interoperability (SOA), flexibility and variability (SPL). Our interest in undertaking 
this case study lies in specifying 
electricity provision as a service and the Smart Grid as a service product line.

\subsection{The Smart Grid as a Product Line}
The generic Smart Grid will be modeled as a family of products with basic components for basic products 
and specialized properties for some of the products, such as:
\begin{itemize}
\item storage;
\item renewables, varying with weather, time, season and other intermittent effects;
\item load shifting, the practice of managing electricity supply and demand so that peak energy use
is shifted to off-peak periods;
\item vehicle to grid (V2G), establishing a viable transparent business model, 
guaranteeing the availability and controllability of electric vehicles (EV) and V2G capacity as well as
accurate forecasting of renewable energy supply and demand.
\end{itemize}
Load shifting and V2G can reduce the energy storage capacity required to maintain power quality.

\begin{figure}[!h]
\begin{center}
\includegraphics[width=0.8\textwidth]{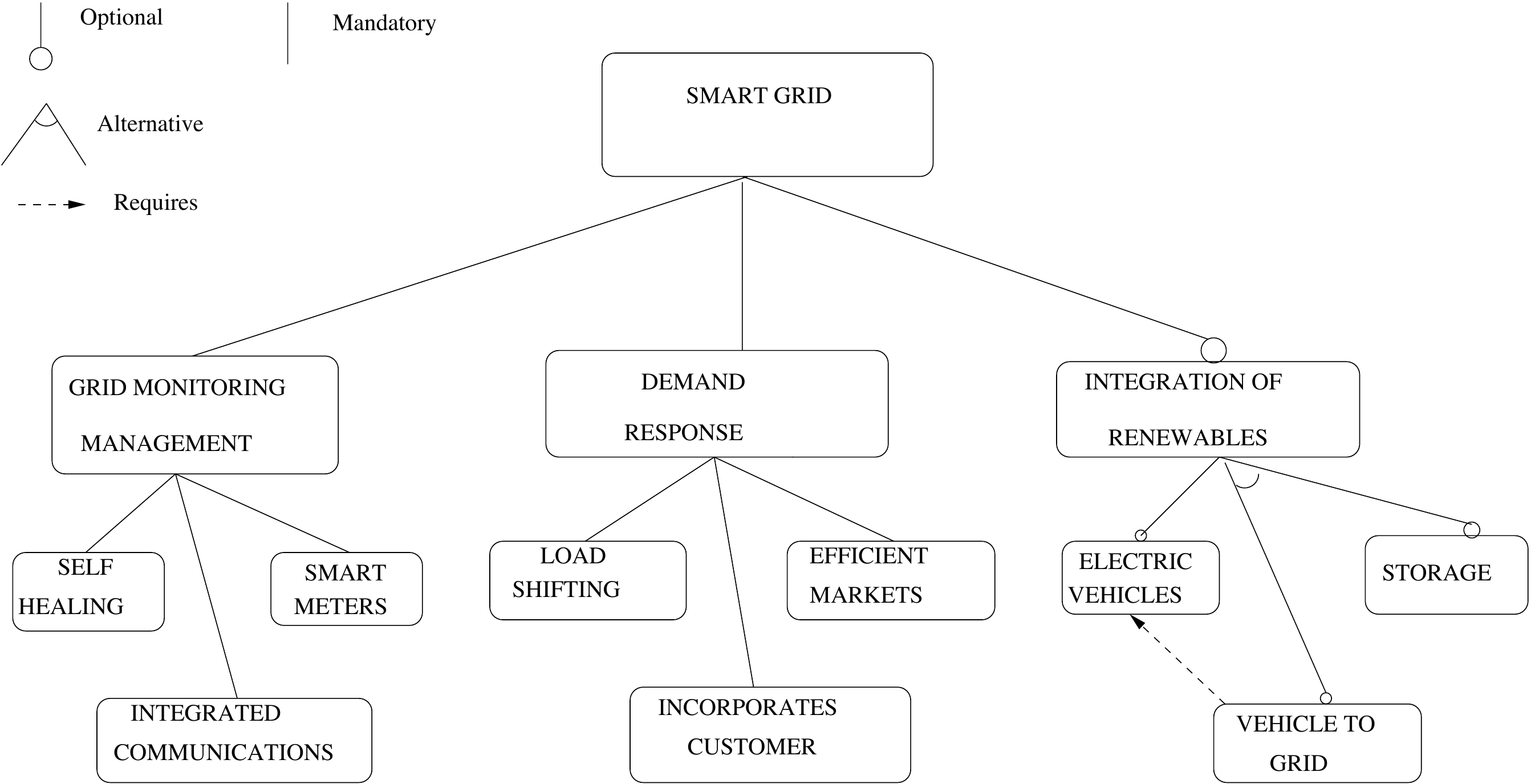}
\caption{Feature model for the Smart Grid family}
\label{Fig:Model} 
\end{center}
\end{figure}

From the Smart Grid family in Figure~\ref{Fig:Model} we can develop up to four different products, all the 
while utilizing the basic 
architecture. Similarly, given an MTS model of the Smart Grid family, we can use Definition~\ref{def:MTS product derivation} to derive  products.

\begin{figure}[!h]
\begin{center}
\includegraphics[width=0.55\textwidth]{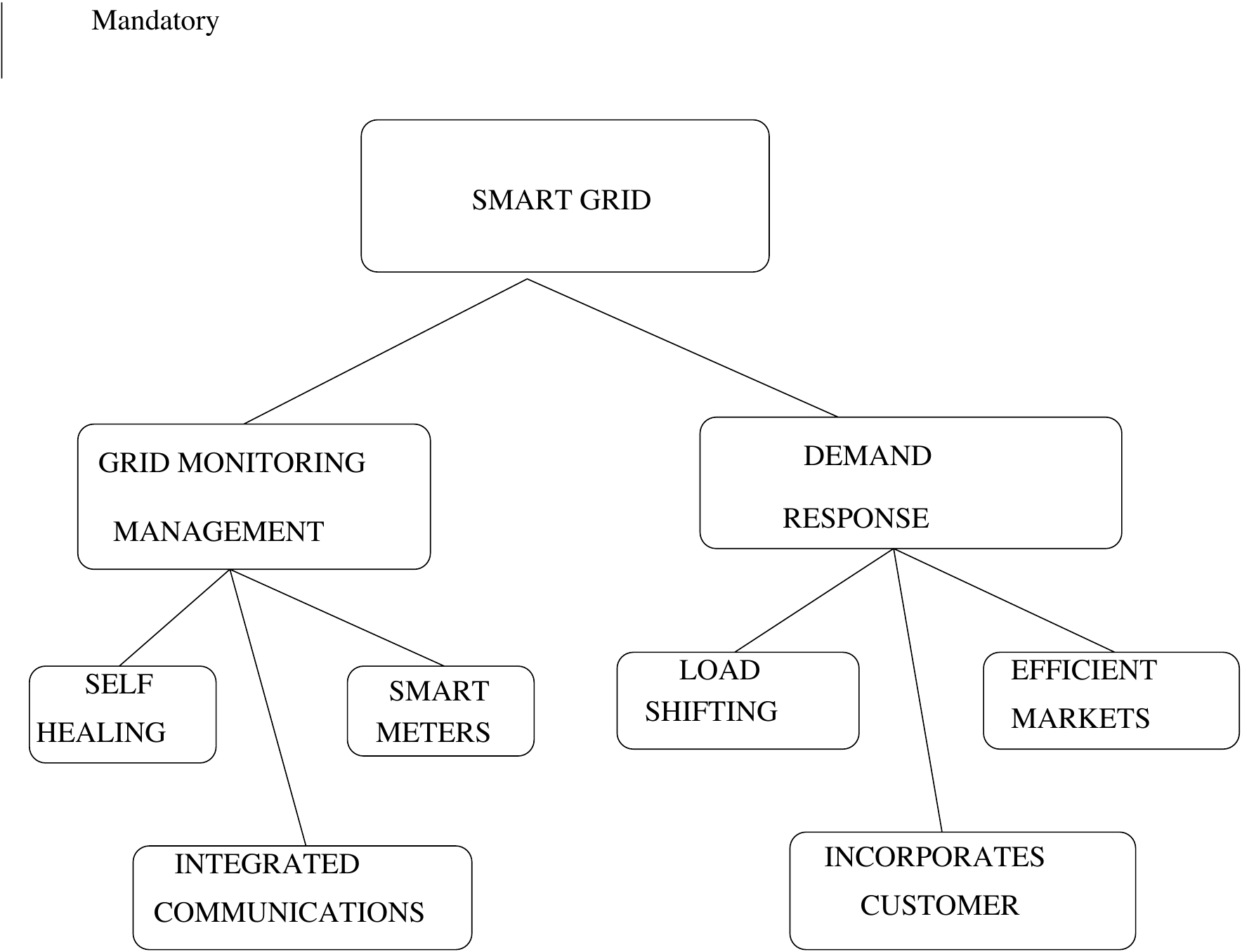}
\caption{Feature model for a product without integration of renewables}
\label{Fig:Mod} 
\end{center}
\end{figure}

One of the most obvious ones is a product without the integration of renewables, shown in 
Figure~\ref{Fig:Mod}, which represents 
most of the existing electricity grids today and in which all the features are mandatory.

This product contains a Demand Response component, referring to dynamic demand mechanisms to manage 
customer consumption of electricity in response to supply and the programs to achieve that goal. 
 
From the utility company point of view, there is a virtual power plant 
where the supply of electricity is managed. In place are technologies that allow the utility to talk to 
devices inside the customer premise. They include such things as load control devices, smart thermostats 
and home energy consoles for sensing, so as to provide information to consumers and operators so 
that they better understand consumption patterns and make informed decisions for more effective 
use of energy~\cite{DR}. 

These are essential to allow customers to reduce or shift their power use during 
peak demand periods. Demand response solutions play a key role in several areas: pricing, emergency 
response, grid reliability, infrastructure planning and design, operations and deferral. 

\begin{figure}[h!]
\begin{center}
\includegraphics[width=0.7\textwidth]{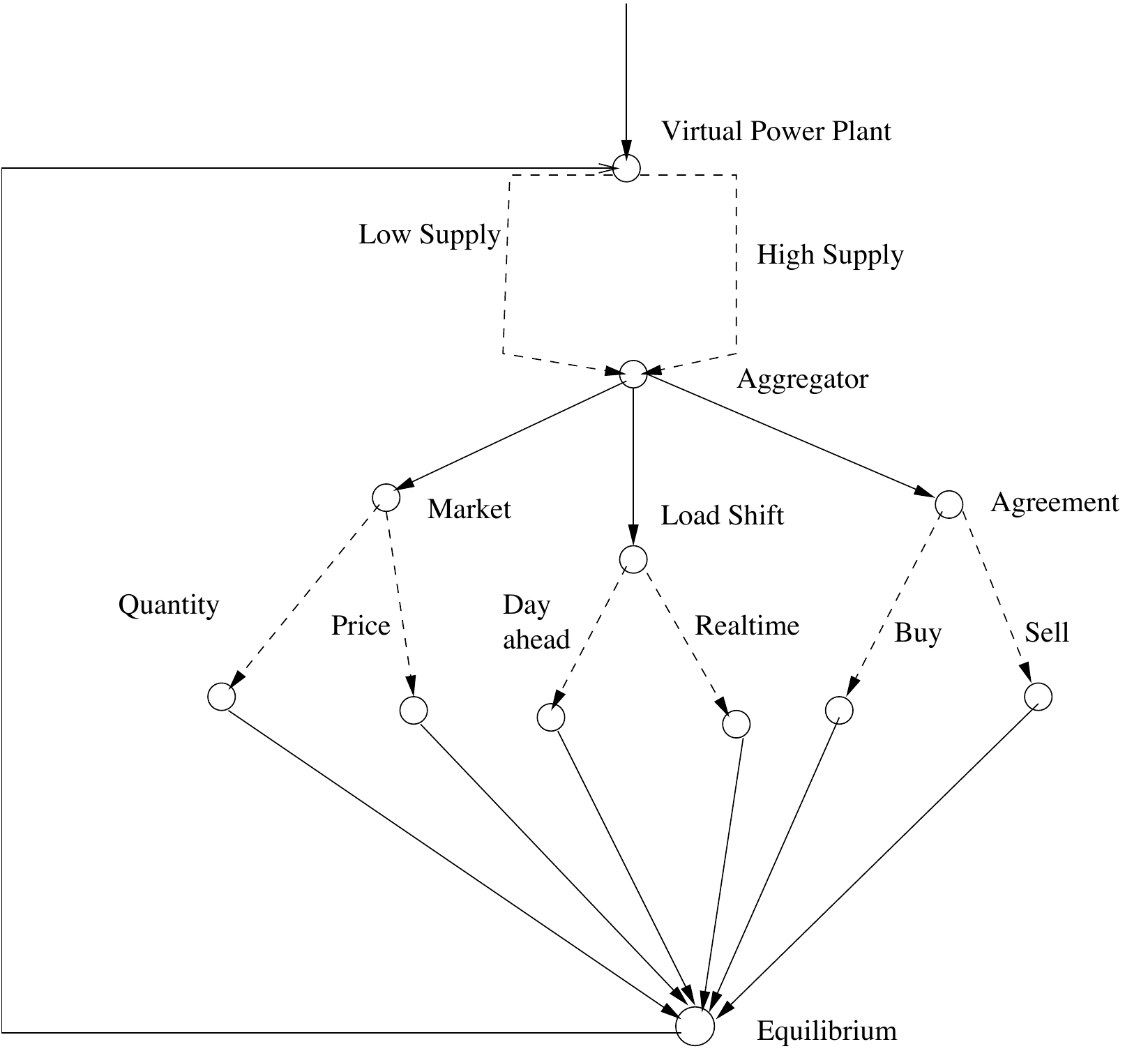}
\caption{MTS for the demand response function}
\label{fig:Model1} 
\end{center}
\end{figure}

The part of the MTS model of the Smart Grid family which is relevant for this component is shown in Figure~\ref{fig:Model1}.

\begin{figure}[h!]
\begin{center}
\includegraphics[width=0.6\textwidth]{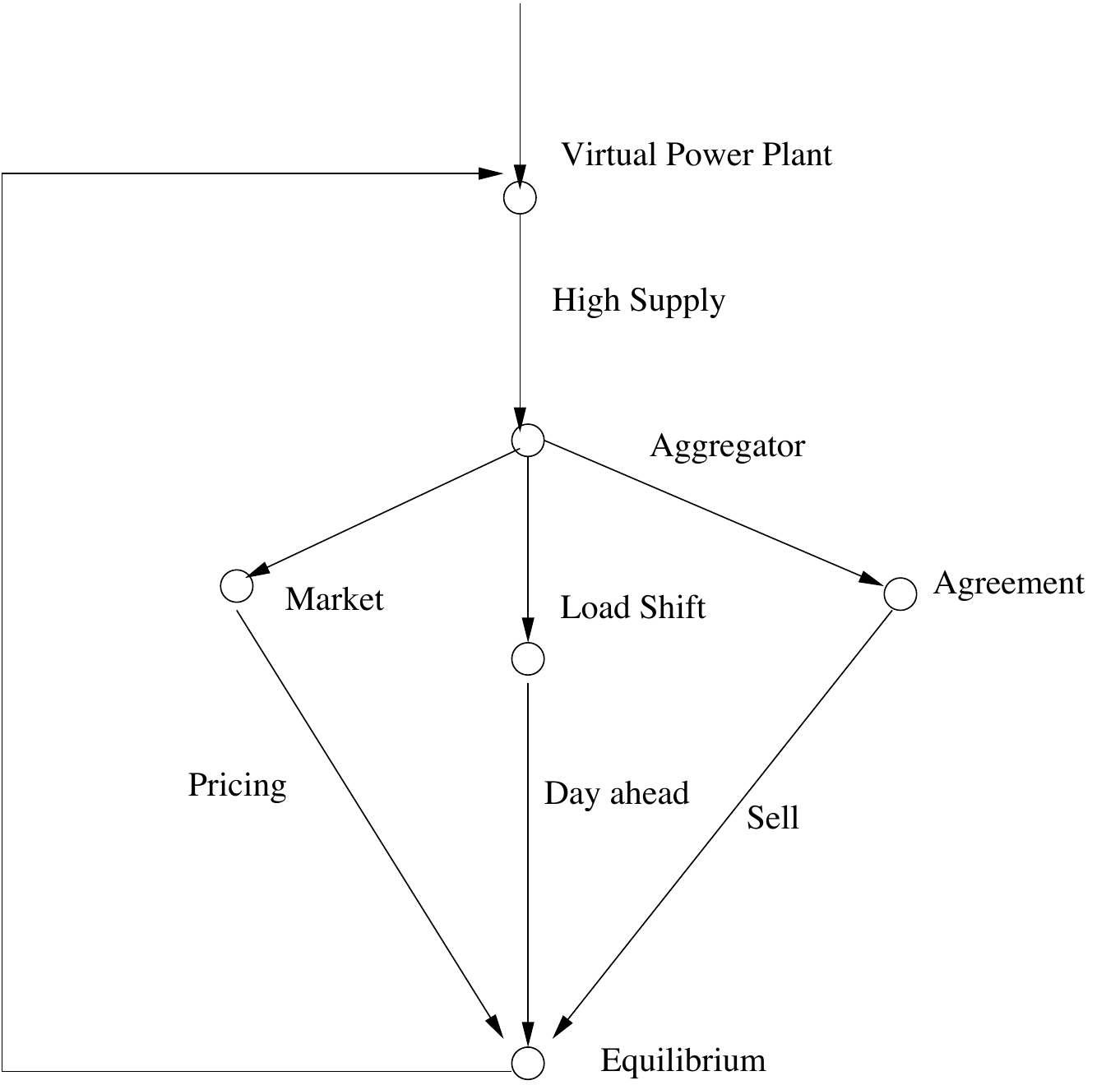}
\caption{A behavioural description of demand response when supply is high}
\label{fig:Model2} 
\end{center}
\end{figure}

\noindent
Thus, the three branches stemming out of the aggregator perform the following functions:
\begin{enumerate}
\item Offer flexible tariffs including critical peak pricing and real-time pricing.
\item Two-way communications allow for pricing information to be transmitted to customers based on 
price changes each day and at timed intervals, determined by software at the enterprise level to allow 
real time or day ahead management.
\item Exception pricing as well as price changes associated with system emergency conditions 
and quantity available to enable the customer to either buy or sell depending on their capacity.
\end{enumerate}
From this we can break down behaviour as shown in Figure~\ref{fig:Model2}, highlighting the behaviour of the 
system when supply of electricity is high, represented as:
$$\text{DRH} := \{\text{High Supply}, \text{Agreement}, \text{Sell}, \text{Equilibrium}\}$$
This means that when the utility has excess supply of electricity, it will take advantage of 
existing agreements with their customers to sell and allow the system to get back to a state of 
equilibrium.

\subsection{Encoding the Product Line in Orc} 
The product demand response is realized in terms of service orchestration using the combinators as follows. From Figure~\ref{fig:Model1}, we model the two branches on the right. We need to spawn two independent 
threads at a point in the computation in this case depending on whether we are load shifting or 
executing an existing agreement and resume 
the computation after both threads at the equilibrium point. Therefore, we call sites 
$\Ind{\textit{real\_time}}{\textit{day\_ahead}}$ and $\Ind{\textit{sell}}{\textit{buy}}$ in parallel and compose using the 
asymmetric parallel combinator. The call then publishes the values as a tuple after they both complete 
their executions.
$$DR:= {{\bf let}\ (u,v)}<{Load\_shift}<{\Ind{real\_time}{day\_ahead}}<{Agreement}<{\Ind{sell}{buy}}$$
The values published by this expression are the values contained in site {\bf let}, which acts as a container 
for the first result published and releases both when the second value is received.

In the same way we can model the fact that these two features are alternatives. This means that once we 
instantiate the computation \textit{Load\_shift\/} we are not in a position to execute \textit{Agreement\/} 
and vice versa. We therefore have
$$(\textit{if}(f)\gg L\_s) \mid (\textit{if} (\neg f)) \gg A \ll f \in (\Ind{\textit{real\_time}}{\textit{day\_ahead}} \gg {\bf let}(\textit{true})) \mid (\Ind{\textit{buy}}{\textit{sell}} \gg {\bf let}(\textit{false}))$$
in which, due to restrictions in space, we have used the following abbreviations: $f$ for $\textit{flag}$ (our container for the
computation published first), $L\_s$ for $\textit{Load\_shift}$ and $A$ for $\textit{Agreement}$. As a start, we 
call $L\_s$ and $A$ in parallel by utilizing the independent parallel combinator and await for values published 
by either $\Ind{\textit{real\_time}}{\textit{day\_ahead}}$ or $\Ind{\textit{sell}}{\textit{buy}}$ to determine which of the computations 
is executed. The value published first by the asymmetric parallel combinator is held. The site $\textit{if}$ 
returns the value held in $A$ if $A$ is true, and otherwise does not respond and thus, completes the 
alternative service reasoning.

\paragraph{Why Orc?\/} The dynamic nature of this illustration highlights the dynamic nature of the various 
components and services of the Smart Grid, which is the main reason why we opted to 
work on the Orc calculus. Dynamic service, market management and pricing are basic building blocks of 
a Smart Grid system while
Orc allows for the dynamic combination of services and the dynamic reconfiguration of 
software systems. The idea of invoking a published service instead of developing an isolated function 
leads a revolution of application development~\cite{Jifeng}. Thus, electricity business jobs can be arranged 
by orchestrating the 
services which can provide computation resources or functional support.

\subsection{Semantics}
The semantics is operational, asynchronous, and based on LTSs~\cite{task-orch}.
We show what this means for our case study.
We have $?k$, which denotes an instance
of a site call that has not yet returned a value, where $k$ is a unique handle that
identifies the call instance. 
The transition relation $A\xrightarrow{a} A$, states that expression
$A$ may transition with event $a$ to expression $A$. There are four kinds of events,
which we call base events:
$$a,b\in \textit{BaseEvent}~::=~!v \mid \tau \mid M_k(v)\mid k?v$$
A publication event, $!v$, publishes a value $v$ from an expression. As is traditional, 
$\tau$ denotes an internal event. The remaining two events, the site call event
$M_k(v)$ and the response event $k?v$, are discussed below.

A site call $M (v)$, in which $v$ is a value, transitions
to $?k$ with event $M_k(v)$. The handle $k$ connects a site call to a site return --- a
fresh handle is created for each call to identify that call instance. The resulting
expression, $?k$, represents a process that is blocked waiting for the response from
the call. A site call occurs only when its parameters are values; in $M(x)$, in which
$x$ is a variable, the call is blocked until $x$ is defined.

The composition rules are straightforward, except in some cases in which subexpressions 
publish values. 

Consider the independent parallel combinator $\Ind{A}{B}$. When $A$ publishes 
a value ($A\xrightarrow{!v} A$ ), it creates a new 
instance of the right-hand side, $[v/x].B$, the expression in which all free
occurrences of $x$ in $B$ are replaced by $v$. The publication $!v$ is hidden, and the
entire expression performs a $\tau$ action. Note that $A$ and all instances of $B$ are
executed in parallel. Since the semantics is asynchronous, there is no guarantee 
that the values published by the first instance will precede the values of
later instances. Instead, the values produced by all instances of $B$ are interleaved
arbitrarily.

Asymmetric parallel composition, $\Par{A}{x}{B}$, is similar to parallel composition, except
when $B$ publishes a value $v$. In this case, $B$ terminates and $x$ is
bound to $v$ in $A$. One subtlety of these rules is that $A$ may contain both active
and blocked subprocesses --- any site call that uses $x$ is blocked until $B$ publishes.

We now check one of our expressions:
$$DR := {{\bf let}\ (u,v)}<{\textit{Load\_shift}}<{\Ind{\textit{real\_time}}{\textit{day\_ahead}}}<{\textit{Agreement}}<{\Ind{\textit{sell}}{\textit{buy}}}$$
An execution of $DR$ is as follows: the first step $\Ind{\textit{sell}}{\textit{buy}}$ publishes $\textit{sell}$ or $\textit{buy}$, which is bound 
to $\textit{Agreement}$:
$$[\textit{sell}/\textit{Agreement}].\Ind{\textit{sell}}{\textit{buy}}$$ 
or 
$$[\textit{buy}/\textit{Agreement}].\Ind{\textit{sell}}{\textit{buy}}$$
The same strategy is employed for the next step and we have an event $\tau$ due to the site $\textit{let}$ as it 
receives one value before the other.

\section{Conclusion and Future work}\label{FW}
Correct modeling of service product lines appears to be important --- if not vital --- in order for them to 
provide the same level of accuracy and support as their earlier counterparts, service-oriented applications 
and SPLs. We have seen that Orc, being a language for orchestration at a moderately 
abstract level, provides important support for this. We have proposed that services can be modeled in a 
new way by incorporating variability notions from SPLs. It is perhaps unexpected how 
direct the resulting reasoning is to variability in product family descriptions using feature modeling. 
This is due to the extent to which Orc captures the sequences of publications. 

We have not fully formalized 
service product lines, but the techniques and tools identified and the 
relationships established amongst them are a firm foundation for this. We plan to extend Orc in order 
to support the abstract layer provided by the MTSs and to allow verification by means of the {\sc vaCTL} logic. Feature 
models will remain our low-level representation of our service product lines. To fully formalize and 
verify service product lines, we require that Orc has the capability to 
integrate the tools. We also intend to extend the LTS-based semantics of Orc to an MTS-based semantics in 
order to utilize the {\sc vaCTL} logic for verification.

\section*{Acknowledgments}

We thank the referees and participants of WWV 2011 for their useful comments before, during and after the workshop, which have considerably improved this paper.

\nocite{*}
\bibliographystyle{eptcs}
\bibliography{Proposal}

\end{document}